\documentclass[twocolumn,prl]{revtex4}

\usepackage{amsmath}
\usepackage{amssymb}

\begin{document}
\title{Comment on ``NMR Experiment Factors Numbers with Gauss Sums''}
\author{J.~A. Jones}
\email{jonathan.jones@qubit.org} \affiliation{Centre for Quantum
Computation, Clarendon Laboratory, University of Oxford, Parks
Road, OX1 3PU, United Kingdom}


\noindent\textbf{Comment on ``NMR Experiment Factors \\Numbers with Gauss Sums''}

Mehring \textit{et al.} have recently described an elegant nuclear magnetic resonance (NMR) experiment \cite{Mehring} implementing an algorithm to factor numbers based
on the properties of Gauss sums.  Similar experiments have also been described by Mahesh \textit{et al.}~\cite{Suter}.  In fact these algorithms do not factor numbers
directly, but rather check whether a trial integer $\ell$ is a factor of a given integer $N$.  Here I show that these NMR schemes cannot be used for factor checking
without first implicitly determining whether or not $\ell$ is a factor of $N$.

The method is based on a property of truncated Gauss sums
\begin{equation}
{\cal A}_N^{(M)}(\ell) = \frac{1}{M+1} \sum_{m=0}^M \exp\left[- 2\pi i\, m^2\frac{N}{\ell}\right],
\end{equation}
namely that ${\cal A}_N^{(M)}(\ell)=1$ if $\ell$ is a factor of $N$, and that its magnitude is small otherwise, thus allowing factors to be distinguished from
nonfactors as long as the truncation parameter $M$ is not too small.  Mehring \textit{et al.} in fact evaluate the closely related sum
\begin{equation}
\overline{\cal C}_N^{(M)}(\ell) = \frac{1}{M+1} \sum_{m=0}^M \cos\left(2\pi m^2\frac{N}{\ell}\right)
\end{equation}
where I have neglected the effects of relaxation, which is included in their treatment but is not critical to this discussion.  They achieve this by generating a set of
spin echoes by applying a series of $180^\circ$ pulses with phases $\phi_k$ given by
\begin{equation}
\phi_k= \left\{
\begin{array}{cc}
(-1)^k(2k-1) \pi \frac{N}{\ell} &\text{ for } k \ge 1\\[2mm]
0&\text{ for } k =0.
\end{array}\right.\label{eq:phik}
\end{equation}
They then proceed to demonstrate an experimental realization of the algorithm for $N=157573$ and to discuss a numerical simulation of the algorithm for the 24 digit
number $N=1062885837863046188098307$.

While this algorithm does in fact work, it cannot be used in any useful way.  The success of the algorithm relies on the initial calculation of values of $\phi_k$; this
is equivalent to the evaluation of the ratio $N/\ell$, and any calculation of $\phi_k$ must be performed with sufficient precision to indicate whether or not this ratio
is an integer, and thus whether or not $\ell$ is a factor of $N$.  Similar comments apply to the methods of Mahesh \textit{et al}.

Mehring \textit{et al.} then suggest that their method can perhaps be extended using Liouville space quantum computing \cite{Madi} to provide an efficient factoring
algorithm. It is difficult to comment on this as no details are provided, but it seems highly plausible that similar arguments would apply.

In passing I note that the method of Gauss sums works by determining whether or not the ratio $N/\ell$ is an integer, and this is only useful for identifying factors if
$\ell$ can be confined to the integers. For example, it is easy to distinguish $17$, which is a factor of $157573$, from 18, which is not, by using their corresponding
Gauss sums.  However the trial number $157573/9268\approx17.0018343$ gives an equivalent peak in the Gauss sum although it obviously does not correspond to a factor.

For these reason methods based on Gauss sums will only be useful if is possible to avoid explicit division in the algorithm and the integral nature of $\ell$ is built
directly into the implementation.

I thank Ian Walmsley for helpful discussions.

\vspace{0.5cm}
\noindent J. A. Jones\\
Centre for Quantum Computation\\
Clarendon Laboratory, University of Oxford\\
Parks Road, Oxford OX1 3PU, UK

\end{document}